\documentclass[pra,twocolumn,showpacs,preprintnumbers,amsmath,amssymb]{revtex4-1}

\usepackage{graphicx,amsmath}
\usepackage{dcolumn}
\usepackage{bm}

\begin{document}

\title{Kelvin-Tkachenko waves of few-vortex arrays in trapped Bose-Einstein condensates}
\author{T. P. Simula$^{1,2}$ and K. Machida$^{2}$}
\affiliation{$^1$School of Physics, Monash University,Victoria 3800, Australia\\$^2$Department of Physics, Okayama University, Okayama 700-8530, Japan}

\begin{abstract}
We have calculated the low-lying elementary excitations of three-dimensional few-vortex arrays in trapped Bose-Einstein condensates. The number of different Kelvin-Tkachenko vortex wave branches found matches the number of vortices in the condensate. The lowest odd-parity modes exhibit superfluid gyroscopic vortex motion. Experimentally, these modes could be excited and observed individually or in connection with the formation and decay of quantum turbulence. 
\end{abstract}

\maketitle

\section{Introduction}
Rotation exerted on a superfluid forces topological transmutations in its structure due to the nucleation of quantized vortices \cite{Feynman1955a}. Such remarkable behavior sets quantum liquids including superconductors, superfluid helium and gaseous Bose-Einstein condensates apart from Newtonian fluids. Vortex degrees of freedom introduce additional normal modes in the spectra of superfluid systems altering their dynamical properties. Excitations of multi-vortex systems are pivotal to the understanding of the microscopic mechanisms which govern quantum turbulence in superfluid systems \cite{Vinen2002a,Henn2009a}. Vortex waves are also relevant to the dynamics of quenched quantum phase transitions and such excitations are likely to accompany spontaneously formed vortices in the birth of superfluids \cite{Weiler2008a,Freilich2010a}. Furthermore, the elementary excitations of three-dimensional rotating vortex lattices may play role in post-glitch relaxation phenomena observed in rotating neutron stars whose inner parts are thought to be filled by neutron superfluids \cite{Pines1985a}. 
 
 Quantized vortex lines in simple rotating superfluids arrange themselves into the same triangular lattice structure as predicted for superconductors by Abrikosov \cite{Abrikosov1957a}. Kelvin waves are helical modes which propagate along the vortex line. They have been extensively studied in classical fluids since the seminal work by Thomson (Lord Kelvin) \cite{Thomson1880a}. Their dispersion relation has the characteristic long-wavelength form $\omega_{\rm K}\approx \Gamma k^2 \ln(1/|k|r_c)$, where $\Gamma$ denotes the circulation around the axis of the vortex, $r_c$ is a core parameter, and $k$ is the wave vector. Similar dispersion relation holds also for Kelvin waves of quantized vortex lines in superfluids \cite{Pitaevskii1961a,Simula2008a}. 

Vortex lattices support collective Tkachenko waves in which the individual vortices trace elliptical trajectories around their equilibrium positions in the plane perpendicular to the axis oriented along the vortex lines \cite{Rajagopal1964a,Tkachenko1966a}. In contrast to axial Kelvin waves, transverse Tkachenko vortex waves have thus far only been observed in superfluid systems. The generic dispersion relation for Tkachenko waves in uniform systems, 
$\omega^2_{\rm T} \approx  c^2_s c^2_{\rm T} k^4 / (c^2_s k^2 +4\Omega^2)$,
where $c_s$ and $c_{\rm T}$, respectively, denote the propagation speed of sound and Tkachenko wave, changes its character from linear (stiff limit) to quadratic (soft limit) as the rotation frequency $\Omega$ of the system is increased \cite{Sonin1987a}. Tkachenko waves in helium superfluids were discussed using continuum theories by Fetter and Williams \cite{Fetter1975a,Williams1976a}, Sonin \cite{Sonin1976a,Sonin1987a} and Baym and Chandler \cite{Baym1983a}. The stiff modes were also detected in experiments \cite{Andereck1980a}.

Observations of vortices and their dynamical properties become particularly transparent in weakly interacting quantum gases. Orbital motion of a single quantum vortex was imaged in a Bose-Einstein condensate of sodium gas by Anderson \emph{et al.} \cite{Anderson2000a}. Such vortex motion is a manifestation of the fundamental Kelvin wave. Kelvons of higher axial wave numbers were created and observed by Bretin \emph{et al.} in elongated Bose-Einstein condensates \cite{Bretin2003a}. Further discussion of those experiments can be found e.g. in Refs. \cite{Mizushima2003a,Fetter2004a,Simula2008a,Simula2008b}. Direct imaging of planar Tkachenko waves in rotating Bose-Einstein condensates were reported by Coddington \emph{et al.} \cite{Coddington2003a}. Baym explained the observed excitation frequencies by calculating the Tkachenko modes in two-dimensions using continuum elasticity theory accounting for the compressibility of the condensate \cite{Baym2003a},  see also Refs. \cite{Cozzini2004a,Sonin2005a,Simula2004a,Mizushima2004c,Baksmaty2004a}.

Here we obtain, for the first time, a complete microscopic description of the lowest excitation modes of few-vortex arrays in harmonically trapped rotating Bose-Einstein condensates by employing the Bogoliubov-de Gennes wavefunction formalism in fully three-dimensional configurations. We explicitly account for both Kelvin and Tkachenko contributions to the collective vortex motion and mode frequencies and include finite-size effects such as harmonic confinement and bending of vortex lines. Together with sound waves, such elementary excitations exhaust the low-lying collective modes in a vortex lattice system. They are also the key to understanding more complex phenomena such as vortex interconnections inherent in quantum turbulence.

\section{Model}
The zero temperature Bose-Einstein condensate ground state wavefunction $\phi({\bf r})$ and its dynamics is modeled remarkably well by the Gross-Pitaevskii equation
\begin{equation}
i\hbar\partial_t \phi({\bf r},t)  = \mathcal{L}({\bf r},t)\phi({\bf r},t) 
\label{GP}
\end{equation}
where
\begin{equation}
\mathcal{L}({\bf r},t) = -\frac{\hbar^2\nabla^2}{2m}+V_{\rm ext}({\bf r})+g|\phi({\bf r})|^2  - \Omega L_z.
\label{GPL}
\end{equation}
In Eq.(\ref{GPL}) the constant $g$ determines the strength of $s-$wave interactions between particles of mass $m$, $L_z$ is the component of the angular momentum operator parallel to the rotation axis and  the external potential $V_{\rm ext}({\bf r})=m [\omega^2_\perp(x^2+y^2) + \omega_z^2z^2]/2$ is fixed by the transverse $\omega_\perp$ and axial $\omega_z$ harmonic frequencies. The norm of the wavefunction, $\int_V|\phi({\bf r})|^2 d{\textbf r}=N$, yields the number of particles $N$ in the system. We elect $\hbar\omega_\perp$ and $a_0=\sqrt{\hbar/m\omega_\perp}$ to be the units of energy and length, respectively. Our computational system is thus specified by the dimensionless parameters $\omega_z=0.2 \;\omega_\perp$, $gN=1000\;\hbar\omega_\perp a_0^3$, and the trap rotation frequency $\Omega$.

The collective modes of the condensate are solutions to the microscopic Bogoliubov-de Gennes equations 
\begin{eqnarray}
\big[\mathcal{L}({\bf r})-\mu\big]u_q({\bf r}) + \mathcal{M}({\bf r})v_q({\bf r})  &=&  E_q u_q({\bf r}) \notag\\
\big[\mathcal{L^*}({\bf r})-\mu\big]v_q({\bf r}) + \mathcal{M^*}({\bf r})u_q({\bf r})  &=& -E_q v_q({\bf r})
\label{BDG}
\end{eqnarray}
where $\mu=\langle \mathcal{L}({\bf r}) \rangle$,  $\mathcal{M}({\bf r})=\phi^2({\bf r})$ and $u_q({\bf r})$ and $v_q({\bf r})$ are the usual quasiparticle amplitudes and $E_q=\hbar\omega_q$ the corresponding energies, where $q$ uniquely labels the eigenstates. We discretize Eqs.(\ref{BDG})  using a finite-element discrete variable representation \cite{Schneider2006a,Simula2008d} and diagonalize the resulting eigenproblem using a parallelized implementation of an iterative Arnoldi process, for details see \cite{PARPACK}. The collective oscillatory motion of the condensate perturbed by the quasiparticle modes may be visualized by the time-dependent probability density $|\tilde{\phi}_q({\bf r,t})|^2$ with
\begin{equation}
\tilde{\phi}_q({\bf r,t}) = \phi({\bf r}) +  p [ u_q({\bf r})e^{-i\omega_q t} + v^*_q({\bf r}) e^{i\omega_q t}]
\label{UV}
\end{equation}
and the number $p$ controls the amplitude of the quasiparticle excitation mode relative to the ground state \cite{EPAPS}.

\section{Rotating ground states}
\begin{figure*}
\includegraphics[width=2\columnwidth]{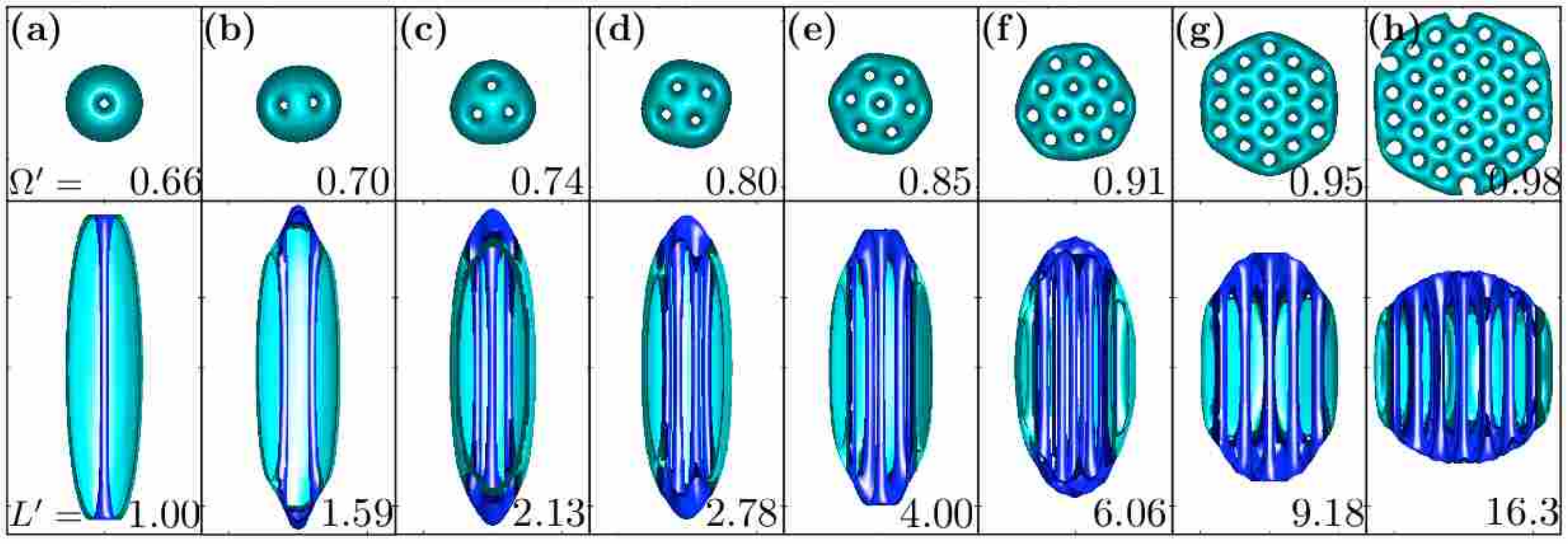}
\caption{(Color online) Rotating ground states for different external angular rotation frequencies $\Omega'$. Frames in the upper row show top $x
$-$y$ view of the condensate density along the rotation axis while the frames in the bottom row display the corresponding side $x$-$z$ view perpendicular to the rotation axis. The frame sides are $14 \;a_0$ and $24\; a_0$ in extent.} 
\label{fig1}
\end{figure*}

Figures \ref{fig1}(a)-(h) display rotating ground states calculated using Eq.(\ref{GP}) for different external angular rotation frequencies of the harmonic trap. Both top view (upper row) and side view (lower row) of the condensate density isosurfaces are shown. Values for the reduced rotation frequency $\Omega'=\Omega/\omega_\perp$ and the angular momentum per particle $L'=\langle L_z \rangle/N\hbar$ are marked in the frames. In the states shown, vortex bending is significant only near the condensate boundaries, however, for $L'<1$ the single vortex state is strongly bent. For our parameters, the five vortex ground state assumes the shape of Olympic rings instead of a five-fold symmetric ring. Six vortex array has a single central vortex and five satellite vortices and seven-vortex state, Fig.~\ref{fig1}(e), completes the full two ``orbital'' lattice. With increasing values of $\Omega'$ the centrifugal effect causes the cloud to expand in the radial direction accompanied with the corresponding axial shortening such that for $\Omega'=0.98$ the shape of the cloud becomes roughly spherical, see Fig.~\ref{fig1}(h). Eventually at high enough rotation frequencies the cloud enters the two-dimensional regime where lowest Landau level physics becomes relevant. 

\section{Collective modes}
\begin{figure}
\includegraphics[width=0.8\columnwidth]{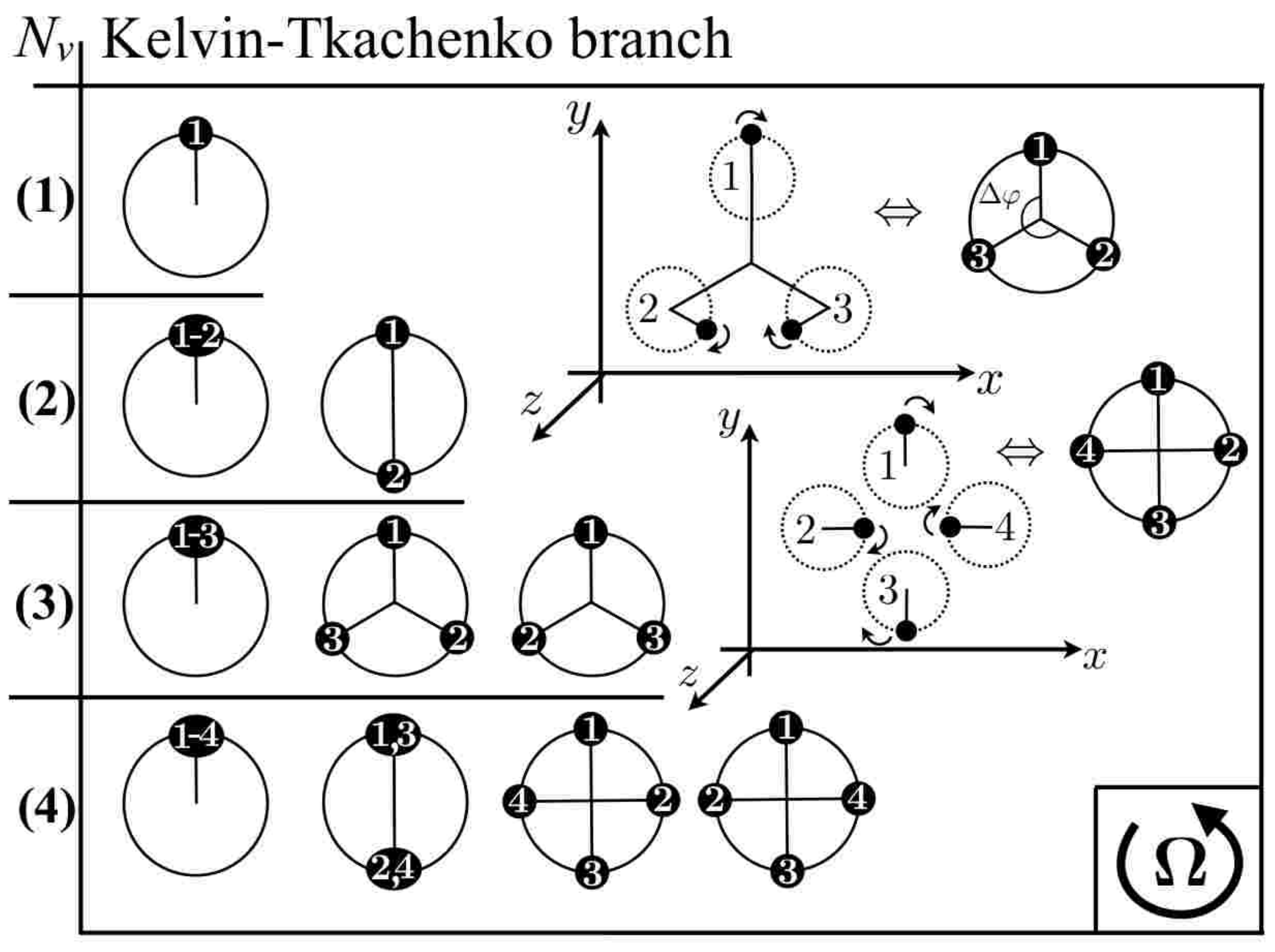}
\caption{Schematic representation of the Kelvin-Tkachenko mode branches. Each diagram corresponds to different vortex wave branch. The two insets clarify the physics encoded in the diagrams. Vortices denoted by the bullets are labelled by numbers in counter-clockwise order. The arrows next to the numbered bullets indicate the direction of local vortex motion.}
\label{fig2}
\end{figure}

When the system is pierced by quantized vortices each added vortex degree of freedom introduces new collective modes in the system. The lowest Kelvin-Tkachenko mode reduces to the fundamental Kelvin mode in the case of a single vortex in the slow rotation limit and to the planar Tkachenko mode in the rapidly rotating (two-dimensional) system. Figure \ref{fig2} diagrammatizes the different vortex wave branches for few-vortex arrays comprising of one through four vortices. Diagrams on rows (1)-(4) apply to the respective ground states shown in Fig.~\ref{fig1}(a)-(d) and the columns correspond to different Kelvin-Tkachenko mode branches, the number of which matches the number of vortices in the system. The bullets mark the relative orientation of the vortices with respect to their equilibrium position such that the numbers inside them enumerate the vortices in the counter-clockwise order. The two insets in Fig.~\ref{fig2} exemplify the interpretation of the diagrams by displaying the locations and directions of motion of the vortices in the $x-y$ plane. Different axial wave numbers $n=0,1,2\ldots$ are implicitly included in each diagram. All modes in the first column describe in-phase oscillations of the vortex array as a whole while the last diagram of each row corresponds to the torsional out-of-phase Kelvin-Tkachenko modes. The diagrams in Fig.~\ref{fig2} represent a qualitatively correct idealization of the actual vortex motion in which the vortices rotate along non-circular trajectories in the direction indicated by the arrows \cite{EPAPS}. Denoting the displacement phase difference between any two neighboring vortices by $\Delta \varphi$, the $N_v$ distinct vortex wave branches satisfy $\Delta \varphi =\{0,1,\cdots,N_v-1\}\frac{2\pi}{N_v}$, where $N_v$ is the number of vortices in the single orbital array.

\subsection{Single vortex}
In the univortex case the sole vortex wave branch corresponds to the Kelvin waves of the quantized vortex \cite{Simula2008a}. The lowest even parity $n=0$ quasiparticle excitation is the fundamental Kelvin mode which manifests itself as an orbital motion (frequently referred to as vortex precession in the literature) of the vortex line around its equilibrium position \cite{Isoshima1999b,Anderson2000a,Svidzinsky2000a,Virtanen2001a}. The lowest odd parity $n=1$ Kelvin mode corresponds to the gyroscopic precession of a tilted vortex with a $\pi$ phase difference between the ends of the vortex line \cite{EPAPS}. Such gyroscopic vortex dynamics was experimentally observed by Haljan  \emph{et al.} \cite{Haljan2001a} and Hodby \emph{et al.} \cite{Hodby2003a}, see also discussion by Stringari \cite{Stringari2001a}. Modes with higher wave numbers $n$ trace the rest of the Kelvin wave spectrum with increasing frequencies \cite{Simula2008a}. An external rotation applied to the system stabilizes these Kelvin modes \cite{Isoshima1999a,Simula2002a}. As a result, all excitation energies are positive with respect to the condensate energy and therefore such Kelvin modes rotate in the opposite sense with respect to the superflow of the condensate. If the stabilizing rotating drive is stopped, some of the lowest modes are shifted to negative energies and therefore those modes change their sense of rotation. The Kelvin modes of an isolated vortex in a trapped Bose-Einstein condensate can be excited via resonant coupling to quadrupole mode as detailed in Refs. \cite{Bretin2003a,Mizushima2003a,Simula2008a,Simula2008b}. 

\subsection{Double vortex}
With two vortices in the condensate their dynamics is enriched by the vortex-vortex interaction. There are two distinct Kelvin-Tkachenko branches corresponding to the two diagrams shown in the row (2) in Fig.~\ref{fig2}. The first one comprises the ``acoustic'' modes which exhibit an in-phase, $\Delta \varphi =0$, elliptically polarized motion of the two vortices around their respective equilibrium positions. The second branch is formed by the ``optical" modes in which the orbital motion of the vortex pair is out-of-phase with a phase difference $\Delta \varphi = \pi$. In the latter radial motion of the vortices is strongly suppressed due to the angular momentum conservation. Consequently the dynamics appears as an azimuthal torsional vibration \cite{EPAPS}. Interestingly, for low momenta the modes in the optical branch have slightly lower excitation energies than the acoustic ones. 

\subsection{Triple vortex}
In the three vortex ground state the vortices arrange themselves into the elementary triangular unit cell of the Abrikosov vortex lattice as shown in Fig.~\ref{fig1}(c). Corresponding to the three vortex degrees of freedom, there are three branches of vortex waves \cite{EPAPS}. In the $n=0$ mode of the first branch, $\Delta \varphi=0$, all three vortices travel in unison around their respective equilibrium positions roughly preserving their triangular relative orientation. The first odd mode $n=1$ of this branch describes gyroscopic vortex motion of the three vortex array as a whole. Such tilting modes were observed experimentally by Smith \emph{et. al} \cite{Smith2004a}. The schematic in the upper inset of Fig.~\ref{fig2} illustrates the motion of the three vortices in the second branch with $\Delta \varphi=4\pi/3$. This kind of three-vortex oscillatory motion was observed by Yarmchuck and Packard who produced photographic movies of few-vortex arrays in superfluid helium \cite{Yarmchuck1981a}. The third branch is comprised of out-of-phase, $\Delta \varphi=2\pi/3$, torsional Kelvin-Tkachenko modes. As for the case of two vortices this last Kelvin-Tkachenko branch, referred to as twisting modes by Chevy \cite{Chevy2006a}, contains the mode with lowest oscillation frequency.
 
\subsection{Quadruple vortex}
Four vortex array supports four branches of Kelvin-Tkachenko vortex waves \cite{EPAPS}. There are an in-phase $\Delta \varphi = 0$ and out-of-phase $\Delta \varphi=\pi/2$ branches respectively corresponding to the first and last columns of the row (4) in Fig.~\ref{fig2}. The motion in the second branch, $\Delta \varphi=\pi$, may be viewed as that of a pair of vortex molecules such that the internal motion of vortices in each molecule is in-phase while the relative motion between the two vortex molecules has a $\pi$ phase difference. Vortex motion in the remaining third branch is illustrated in the lower inset of Fig.~\ref{fig2}.  In it $\Delta \varphi = 3\pi/2$ and the internal motion of each vortex molecule is out-of-phase and also the relative motion between these two vortex molecules has a $\pi$ phase difference.

\subsection{Spectra}
\begin{figure*}
\includegraphics[width=2\columnwidth]{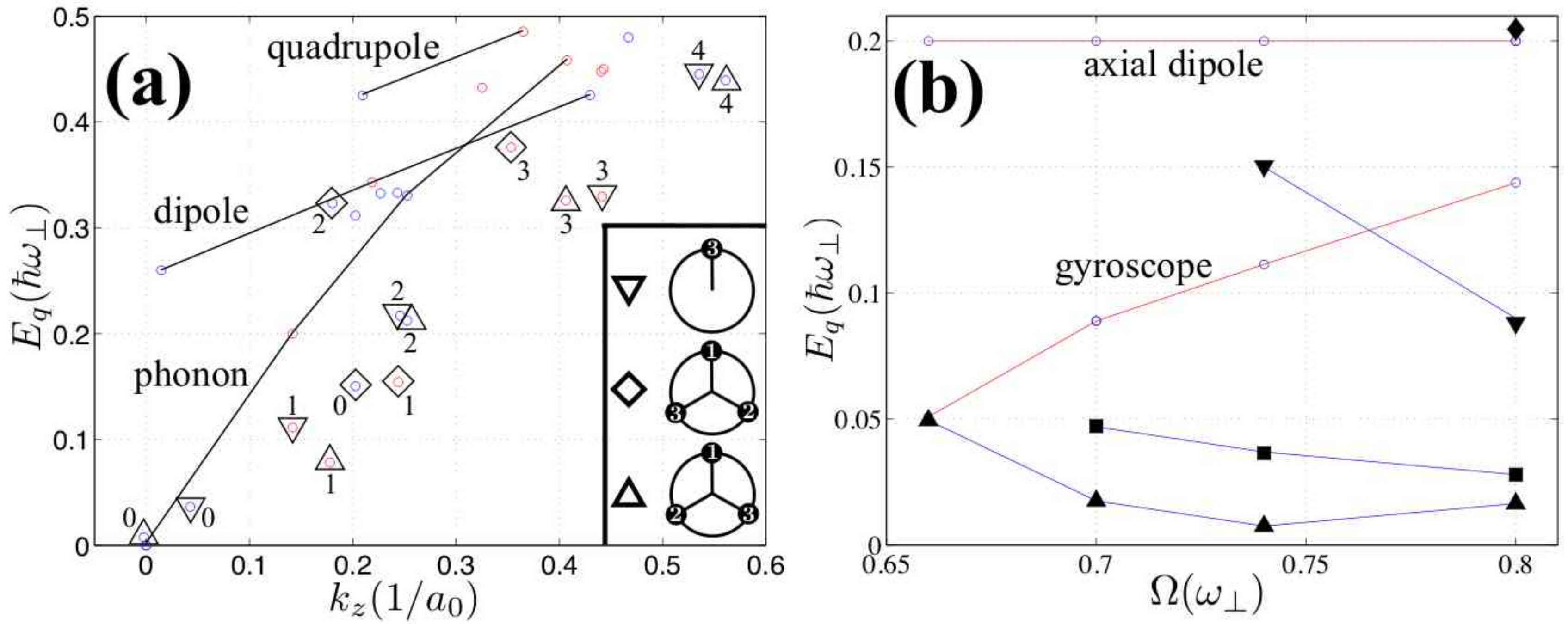}
\caption{(Color online) (a) Spectrum of low-lying quasiparticle states calculated for an array of three vortices with $\Omega'=0.74$. Excitation energies $E_q$ are plotted as functions of their momenta $k_z$ (a number next to a mode denotes its axial wave number $n$). (b) Selected low-lying Kelvin-Tkachenko mode excitation energies $E_q$ as a function of the trap rotation frequency $\Omega$. The straight lines are plotted to guide the eye.}
\label{fig4}
\end{figure*}

Figure \ref{fig4} (a) shows the energies of the lowest quasiparticle excitation modes for the three-vortex ground state of Fig.~\ref{fig1} as functions of their axial momenta defined by $k_z = \sqrt{\langle u_q({\bf r}) | -\partial^2_z | u_q({\bf r}) \rangle+\langle v_q({\bf r}) | -\partial^2_z | v_q({\bf r}) \rangle}$. The three different Kelvin-Tkachenko branches are labeled using the markers indicated in the figure. The lowest odd-parity modes of each branch correspond to the tilted gyroscopic motion of the vortices. Above the Kelvin-Tkachenko branches in energy lie the sound waves with their characteristic linear dispersion relation which becomes curved due to the density inhomogeneity. The lowest co-rotating centre-of-mass mode has the frequency $\omega_\perp-\Omega$ and at $\Omega=\omega_\perp$ it grows without a bound causing the whole condensate to spiral out of the trap due to the centrifugal weakening of the effective trapping potential. 

In Fig. \ref{fig4}(b) we have plotted the $n=0$ modes of each Kelvin-Tkachenko branch with markers $\blacktriangle$, $\blacksquare$, $\blacktriangledown$,  and $\blacklozenge$ corresponding to the diagrams in the last, second but last, third but last  and fourth but last column of each row of Fig.\ref{fig2}, respectively. These modes tend toward the zero point energy as $\Omega$ is increased becoming the constituents of the lowest Landau level in the rapid rotation limit. In addition, the gyroscope modes corresponding to the first odd parity $n=1$ states in the first column of Fig. \ref{fig2} are plotted. The faster the cloud rotates the more energy (torque) is required to tilt the axis of rotation of the system reflecting the fundamental feature (conservation of orbital angular momentum) of gyroscopes to maintain their orientation with respect to the distant stars. The lowest axial dipole modes whose frequencies at 0.2 $\hbar\omega_\perp$ are unaffected by the rotation are joined by the horizontal line. 

\section{Conclusions}
To conclude, we have computed the low-lying elementary excitation modes of three-dimensional few-vortex arrays in rotating trapped Bose-Einstein condensates. The lowest energy Kelvin-Tkachenko excitations correspond to an oscillating torsional motion of the vortex array in the plane perpendicular to the rotation axis. The lowest odd axial modes exhibit superfluid gyroscopic vortex dynamics. The Kelvin-Tkachenko waves could be created in ultra-cold atomic gases experimentally using a suitable combination of laser potentials \cite{Bretin2003a,Coddington2003a} or they might become spontaneously excited in quenched phase transitions \cite{Weiler2008a,Freilich2010a}. Continuous imaging methods should allow experimental investigation of the dynamics of these vortex waves \cite{Freilich2010a}. In superfluid helium vibrating wire experiments \cite{Goto2008a,Blazkova2009a} might enable experimental investigation of such elementary excitations. 

In multi-vortex systems vortex-vortex interactions enrich the internal dynamics of the superfluid. As in the case of a single vortex \cite{Simula2008a}, time-dependent Gross-Pitaevskii simulations show that certain Kelvin-Tkachenko waves of multi-vortex arrays can be excited by resonantly populating the counter-rotating quadrupole mode which couples to these vortex waves. Resonant amplification of multi-vortex Kelvin-Tkachenko waves can lead to vortex reconnections when the amplitude of the modes exceeds the intervortex spacing, onsetting quantum turbulence in these superfluid systems. In future, systematic investigation of interaction dynamics of vortex waves in Bose-Einstein condensates may shed light on the microscopic mechanisms which govern quantum turbulence such as vortex reconnections, cascades and the role of Kelvin-Tkachenko waves in transport and dissipation of turbulent energy.

\begin{acknowledgments}
This work was supported by the Japan Society for the Promotion of Science (JSPS).
\end{acknowledgments}

\end{document}